\documentclass[]{aa}

\usepackage[varg]{txfonts}
\usepackage{lscape}

\begin{document}


\title{The Identification of 93 Day Periodic Photometric Variability for YSO YLW 16A}
\author{Peter Plavchan$^1$, Tina G\"{u}th$^1$, Nadanai Laohakunakorn$^2$, and J. Rob Parks$^3$}
\institute{1 Infrared Processing and Analysis Center, California Institute of Technology, M/C 100-22, 770 S Wilson Avenue, Pasadena, CA 91125; 2 Trinity College, Cambridge, CB2 1TQ United Kingdom; 3 Georgia State University
}

\date{Received date /
Accepted date }

\abstract{}{Periodic variability in young stellar objects (YSOs) offers indirect evidence for an active dynamical mechanism.  Starspots, accretion, stellar companions, and disk veiling can contribute to the photometric variability of YSOs.  }{As part of an ongoing study of the $\rho$ Oph star forming region, we report the discovery of 92.6 day periodic variations for the Class I YSO YLW 16A, observed over a period of three years.  A SED model was fit to available photometric data for the object. }{We propose a triple-system with an inner binary with a period of 93 days eclipsed by a warped circum-binary disk.  The nature of the secondary is unconstrained and could be stellar or sub-stellar.  We report the discovery of a tertiary companion at a projected separation of $\sim$40 AU that could account for the circum-binary disk warp.  This light curve and model is similar to the model we proposed for WL 4 in previous work.  Understanding these systems may lead to insights about the nature of stellar evolution and planetary formation, and provide valuable benchmarks for future theoretical modeling and near- and mid-infrared synoptic surveys of YSOs.}{}

\keywords{circumstellar matter --- stars: pre-main-sequence --- stars: variables: other}

\maketitle

\section{Introduction}

Star formation involves the gravitational collapse of a massive cloud core. Between this initial collapse, and the final contraction onto the main sequence, the protostar is classified as a young stellar object (YSO). These YSOs have ages of a few million years ($\sim$1--10 Myr), and are characterized by high levels of accretion, ejection, and magnetic activity, as well as photometric variability \citep[]{joy42}.  The evolution of YSOs falls broadly into four stages. Class 0 objects consist of a collapsing cloud core. Class I objects are protostars embedded inside a spherically-symmetric infall envelope. Class II objects, also known as ``classical T Tauri stars,'' contain a stable primordial disk.  The dispersion of the stable disk reveals a Class III object, or diskless ``weak-lined T Tauri star'' \citep[e.g.][]{ada87}.  The spectral energy distribution (SED) of YSOs differ from normal stars by exhibiting an infrared excess as the circumstellar material reprocesses the central radiation. The amount of excess is strongly correlated with the evolutionary stage of the YSO. Due to the strength of the infrared emission, it is natural to study YSOs at these wavelengths.

\indent  It is widely accepted that planetary systems form out of primordial protostellar disks, and because such disks are an essential structure in the evolution of Class II YSOs, the study of YSOs can lend valuable insights into the processes by which planets form \citep[][and references therein]{planetformref1,planetformref2}.  At optical wavelengths, some YSOs are observed to exhibit periodic photometric variability \citep[e.g.,][]{rebull01,covey}. The observed variability is generally attributed to the rotational modulation of large cold spots, hot spots, accretion and disk veiling.  Photometric variability driven by rotational modulation of the proto-star are less pronounced at infrared wavelengths, thus improving sensitivity to variability driven by disk-related processes and the subject of many recent NIR and \textit{Spitzer} Space Telescope studies such as YSOVAR \citep[]{morales11,flaherty,flaherty2,faesi}.  

$\rho$ Ophiuchus ($\rho$ Oph) is a nearby ($\sim$135 pc) star-forming region containing a few hundred such YSOs from a few Solar masses down to the free-floating planet mass regime \citep{mam08,marsh}. In this paper we investigate the YSO in $\rho$ Oph: YLW 16A.  YLW 16A ( = IRAS 16244-2432, 2MASS J16272802-2439335, ISO-Oph 143, IRS 44) is classified as a Class I protostar \citep[e.g.][]{luh99, bar05} which has been a notable subject of a previous study at X-ray wavelengths \citep{gro01, ima01}. \citet{ima01} detected an unusual bright X-ray flare, with a peak luminosity of 1.3$\times 10^{31}$ ergs s$^{-1}$. A 6.4 keV emission line was identified, which was attributed to fluorescence of cold neutral iron in the circumstellar gas. An extended $\sim$3400 AU \citep{bec08} nebulosity has been observed around YLW 16A in the infrared \citep[$H$ and $K_s$ bands;][]{sim87, luc98} and at thermal radio wavelengths \citep{leo91, gir04}. High-resolution HST NICMOS imagery, obtained June 1998, reveals two nonpoint sources separated by 0.5$\arcsec$, with flux ratios of 1.5 at 1.1 $\mu$m and 1.1 at 1.6 $\mu$m \citep{all02}. \citet{bec08} interpret the second source as being due to a reflection from a dusty jet inside a bipolar cavity. However, their conclusions do not rule out the possibility of a binary companion. \citet[]{herczeg} also find a resolved binary with CRIRES/VLT, finding the west component to be 0.69$\pm$0.12 mag fainter at $M$-band, with CO and extended H$_2$ emission, but no CO emission from the east component.  \citet{sim87} reported a variability of $\sim$1.0 mag in the $K_s$ band, over a timescale of approximately six months. Finally, \citet[]{evans09} derive an extinction of A$_V$ = 9.8 mag for YLW 16A from the \textit{Spitzer} c2d survey, which would correspond to a $A_J$$\sim$2.  

\indent \citet{dop05} presented a high S/N near-infrared spectrum of YLW 16A, which is essentially featureless. There is, however a sharp H$_2$ emission line at 2.1218 $\mu$m, characteristic of still accreting YSOs, and weak absorption features including the CO 2-0 bandhead at 2.3 $\mu$m.  In general, Class I spectra tend to exhibit weaker features due to veiling and continuum emission characteristic of a protostar surrounded by a thick envelope \citep{gre00}.  \citet[]{covey} measure a local standard rest velocity of 4.41 km/s, and do not identify evidence for a double-lined spectroscopic binary companion, down to the $\sim$1-2 km/s level, from the structure of the radial velocity cross-correlation function in a single epoch spectrum \citep[]{coveypriv}.

\indent Another YSO located in $\rho$ Oph is WL 4, a Class II object, whose periodic photometric variability was discovered by \citet{pla08a} to be 130.87$\pm0.40$ days. The authors suggest a triple-YSO model, consisting of an inner binary and a third companion further out. A circum-binary disk, tilted with respect to the binary's orbital plane by the gravitational influence of the third companion, eclipses each member of the inner binary in turn. Periodic eclipsing of a binary by a circum-binary disk is also the preferred model of the well-studied system KH-15D \citep[][and references therein]{herbst}, as well as the recently discovered object CHS 7797 in the Orion star-forming region \citep[]{rod1,rod2}.

In this paper, we present the discovery of periodic near-IR photometric variability for YLW 16A.  Our analysis parallels much of the analysis in \citet{pla08a}.  We present our observations in ${\S}$2, and results in ${\S}$3.   In ${\S}$4, we discuss the implications of our observations and proposed model.  This discovery demonstrates that systems like YLW 16A, WL 4, KH-15D and CHS 7797 may be common in multiple star-forming regions, constituting a new class of disk eclipsing YSOs.  These systems will be valuable to study the evolution of circumstellar disks around YSOs, and potentially the formation sites for circumbinary planets \citep[]{carter}.

\section{Observations}

The photometry for the $J$, $H$, and $K_s$ bands were obtained from the Two Micron All-Sky Survey (2MASS) Calibration Point Source Working Database (Cal-PSWDB) \citep{skr06}.  Between 1997 and 2001, 2MASS imaged the entire sky in three near-infrared bands, $J$, $H$, and $K_s$. Hourly observations of 35 different calibration fields were used to calibrate the 2MASS photometry. One such field is located in $\rho$ Oph, 8.5$\arcmin$ wide in R.A. by 60$\arcmin$ long in decl., and centered at (R.A., decl.) = (246.80780$^{\circ}$, -24.68901$^{\circ}$).  A total of 1582 independent observations were made of this field, which contains YLW 16A, as discussed in further detail in \citet{pla08a,par10}.

\begin{figure} 
\begin{center}
\includegraphics[width=0.7\textwidth]{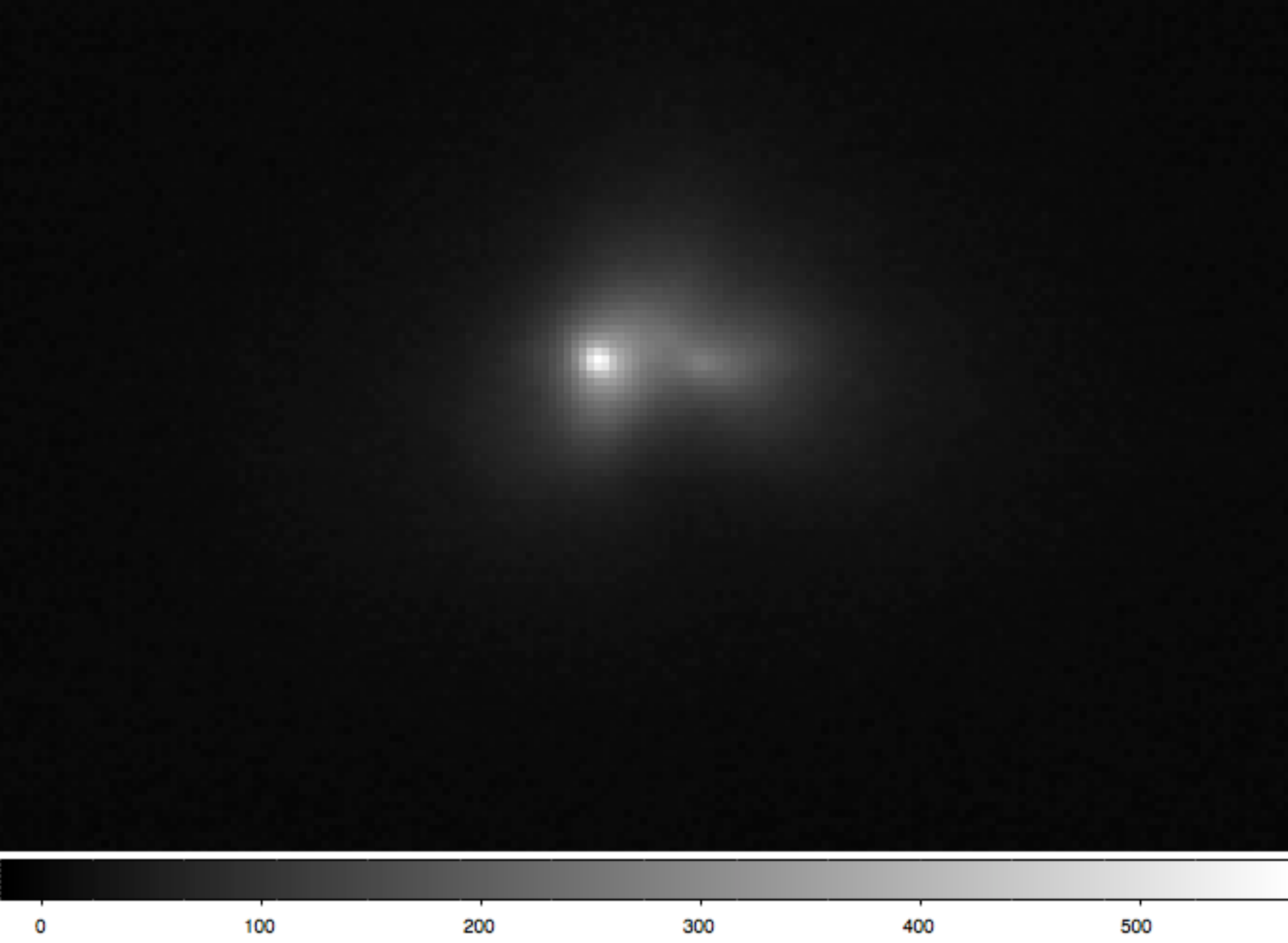}
\includegraphics[width=0.7\textwidth]{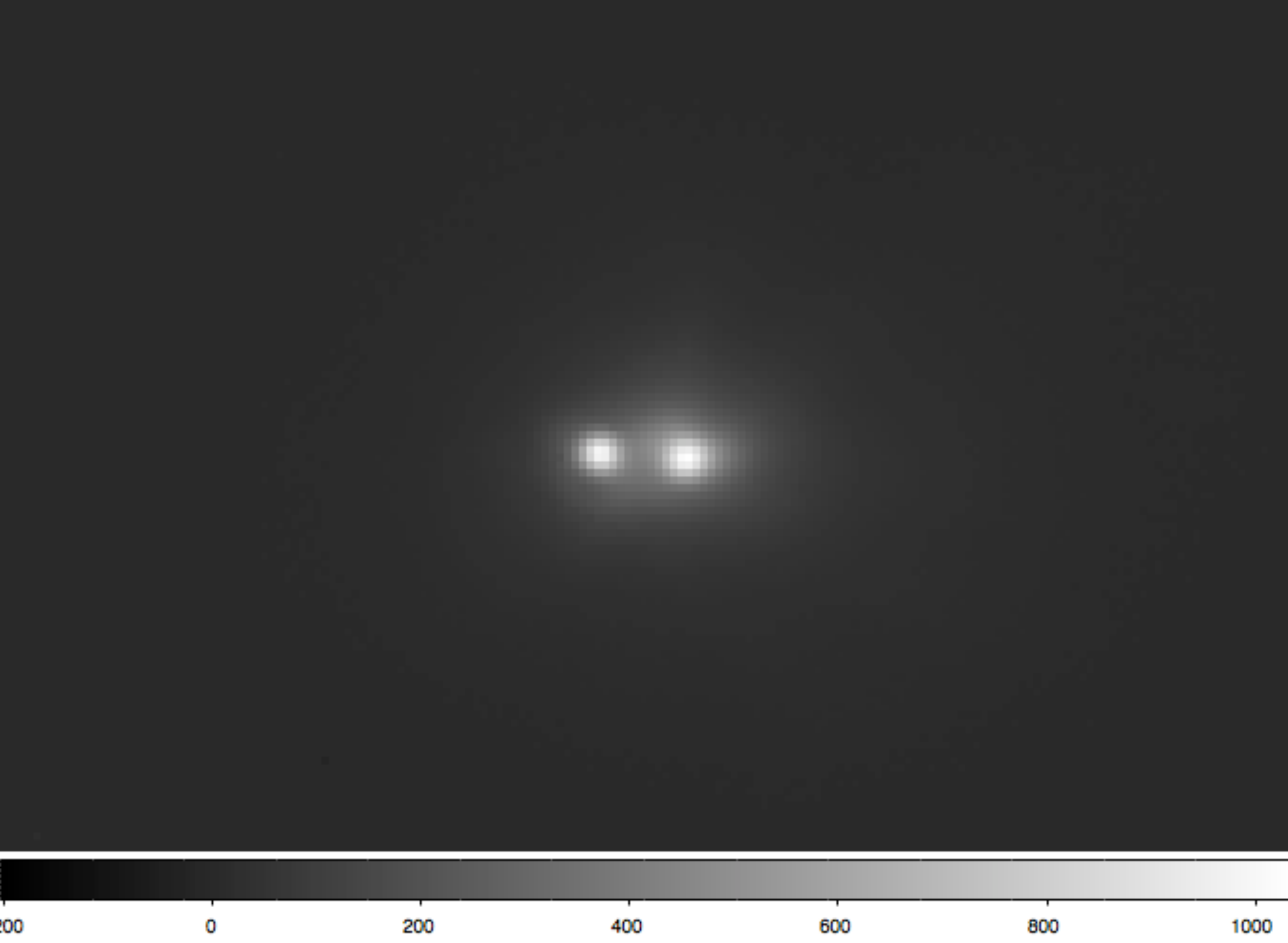}
\end{center}
\caption{NACO imagery of YLW 16A in the $K_s$ (\textit{Upper}) and $L$ (\textit{Lower}) bands. The separation of the two sources in the $L$ band is approximately 0.3$\arcsec$ ($\sim$40 AU projected separation). Both these images were obtained in the faint state ($L$ band phase =0.4656, $K$ band phase =0.6920).  Color bars shown correspond to non-normalized counts. North is up and East is to the left.} 
\label{fig:NACOfig}
\end{figure}

\indent Two NACO images of YLW 16A, taken from the European Southern Observatory (ESO) archive for the instrument, were obtained for the $K_s$ and  $L$ bands \citep{len03, rou03}, as seen in Figure \ref{fig:NACOfig}.  The $L$ band image was obtained on 9 April 2005, while the $K_s$ band was obtained on 30 April 2005.  The NACO images allow us to measure a flux ratio between YLW 16AA and YLW 16AB.  Using aperture photometry, we derive flux ratios from the NACO images of 0.22 and 0.98 at $K_s$ and $L$ respectively.   We calibrate to the total flux from the 2MASS $K_s$ magnitude in the faint state (for the $K_s$ NACO image) of 51.5 mJy and IRAC 3.6 microns in the faint state (for $L$) of 695.8 Jy.

\indent Photometry for YLW 16A was also obtained from the IRAC (3.6, 4.5, 5.8, and 8 $\mu$m) and MIPS (70 $\mu$m) instruments on the Spitzer Space Telescope, as part of the Cores to Disks (c2d) \textit{Spitzer Space Telescope} Legacy program \citep{eva03,pad08}.  YSOVAR has also obtained photometric time-series of YLW 16A at 3.6 and 4.5 $\mu$m during the  \textit{Spitzer Space Telescope} warm mission, which will be part of a separate future publication \citep[]{morales11}.

Finally, photometry was obtained from the literature at 10.8 $\mu$m \citep{bar05}, 850 $\mu$m \citep{jor08}, and 1.2 mm \citep{sta06}.  Excluding the NACO photometry, all other photometry is a blend of the YLW 16A system. The average 2MASS photometry, as well as the additional photometry values are summarized in Table 1.

\section{Analysis and Results}

\subsection{Periodic Variability}

Periodic variability is readily apparent from a visual inspection of the YLW 16A light curve (Figure 2). Using the Lomb-Scargle periodogram \citep{sca82}, the Box Least Squares periodogram \citep[]{bls}, as well as the period-searching algorithm of \citet{pla08b}, a period of 92.62$\pm$0.84 days is identified (also see \citet{par10}).   Access to all three algorithms are available online in interactive form at the NASA Exoplanet Archive, and include rudimentary estimates of the false-alarm probabilities (p-values).  The period and period 1-$\sigma$ error are assessed from the Plavchan periodogram peak and Half-Width Half-Maximum respectively.  All three algorithms detect the signal at 92.6 days, and period aliases of one-half and two times this period.  

The bin-less phase dispersion minimization Plavchan periodogram is adept at detecting arbitrarily-shaped periodic signals, compared to sinusoids for Lomb-Scargle and box-like transits for BLS, and thus detects the 93-day period with higher statistical significance (5.4-$\sigma$).  All three algorithms detect aliased periods at integer fraction multiples of one day (e.g. periods of 1/4,1/3,1/2,2,3, \& 4 days,etc.) with statistical significance that in the case of Lomb-Scargle exceeds the 93-day signal.  This is due to the long-term (e.g. $\sim$400-500 day) variations seen in the light curve aliased with the $\sim$1 observation per day cadence.  Visual inspection of the phased time-series confirms that these are aliased ``false'' periods.   All three algorithms detect these ``red-noise'' long-term trends that are likely astrophysical in origin, but we do not quantify this time-scale further \citep[]{par10}.

Figure \ref{fig:ColorCurves} shows two repeated cycles of the phased light and color curves of YLW 16A.  For JD of 2,450,000.0 (note, not MJD=0), the corresponding phase is 0 in Figure \ref{fig:ColorCurves}.  The average variation between bright and faint states is $\sim$0.5 mag in the $K_s$ band, with a maximum range of $\triangle K_s$ = 0.95 mag.  There is also periodic variability in the $(H-K_s)$ curve.  The shape of the color variations differ from the shape of the photometric variations in the $K_s$ band, with an approximately sinusoidal phased curve with an amplitude of $\sim$0.15 mag and a maximum range of $\triangle$($H-K_s$) = 0.34 mag.  The mean $K_s$ magnitude is 10.22 and the mean ($J-K_s$) color is 7.07.  The low S/N $J$ band data is due to the high extinction towards the system.  There are two data points that appear to be particular errant from the phased time-series in Figure 3 at phases of $\sim$0.7 and 0.8, which may indicate additional sporadic variability such as a flare that was not adequately sampled by the cadence of our observations.

The folded light and color curves show that YLW 16A has both dim and bright states similar to WL 4 as discussed in \citet{pla08a}.  However, the bright state of YLW 16A is about half as long in phase duration as that of WL 4, such that it appears very much like an ``upside-down eclipse''. Additionally, the source is redder in $H-K_s$ when the bright state occurs, which runs counter-intuitive to models for extinction variability \citep[]{par10}.

\subsection{SED Modeling}

A model SED fit is generated, as done in \citet{pla08a}.  Our model SED can include up to three stellar components and five dust components, each with an independent extinction magnitude.  We also allow for a variable broken power law for the system extinction as a function of wavelength (e.g. $A_\lambda\propto \lambda^{\alpha_1,\alpha_2}$ for $\lambda <,> \lambda_0$), and we dynamically color-correct 24 and 70 $\mu$m \textit{Spitzer} photometry. Reddened Phoenix NextGen \citep{hau99} spectra contribute the stellar component(s) of the SED model, while the dust emission is modeled as a blackbody. By varying the temperatures of the stars and dust, as well as their effective radii, the composite curve was obtained using a $\chi^2$ minimization fit to the available photometric data of both the bright and faint states.  Given the multi-parameter fit, we have improved upon the analysis of \citet{pla08a} to include an AMOEBA simplex code parameter optimization \citep[]{metcalf}.  Figure \ref{fig:SEDfit} shows the best model SED fits to each of the bright and faint state SEDs, and the best fit parameters are given in Table 2.  Simplex codes are generally sensitive to the initial guesses.  Thus, our fit likely represents only one of multiple degenerate models that can adequately describe the SED in both the bright and faint states.  However, we have chosen to fix some of the parameters to enforce ensure that these parameters remain consistent between the bright and faint states, and given the limited degrees of freedom. 

\section{Discussion}

From the NACO images in Figure \ref{fig:NACOfig}, we confirm the presence of at least two YSOs in the YLW 16A system.  The $L$ band NACO image shows two sources of approximately equal brightness, whose projected separation is around 0.3$\arcsec$, corresponding to a projected separation of $\sim$40 AU at a distance of 135 pc. The $K_s$ band NACO image shows a more complex nebulosity surrounding the visual binary, indicating at least one component of the binary is deeply embedded in a protostellar envelope. Both NACO images were obtained in the faint state ($L$ band phase =0.4656, $K_s$ band phase =0.6920).  These images do not constrain which of the visual components is the source of the observed photometric variability.  However, if the stellar components are of equal spectral type and radii, the fainter K-band component is a likely candidate given that the system was in a faint state at the time. Time-resolved adaptive optics monitoring is necessary to confirm which component contributes to the photometric variability.

\indent Our SED model fit confirms the Class I nature of YLW 16A.  A fit with two dust blackbody components and no stellar components yields an unphysical extinction power law, and thus the SED model necessitates at least one stellar component.  The large extinction of the stellar components and potential differences in infrared excess do not motivate/yield any useful constraints on multiple component stellar radii nor temperatures, even though the NACO images indicate at least two components.  For example, a higher stellar temperature is partially degenerate with a larger stellar extinction.  Rather, we derive a single composite set of stellar parameters from the fit of $T_*\sim$3500 K and effective stellar radius of $\sim$4.66 R$_\odot$.   If the system consists of three stars with equal radii and temperatures in the bright state, the corresponding stellar radii of each of the three components would be $\sim$2.7 $R_\odot$.  If the system consists of two stars with equal radii in the faint state -- e.g., a third star is completely extincted/eclipsed by a primordial disk -- the corresponding stellar radii of each of the two components would be $\sim$3.0 $R_\odot$.  These radii are plausible for Class I YSOs.  Our value of $A_J\sim10$ yields a visual extinction much larger than that derived in \citet[]{evans09}, despite providing an initial guess for the stellar extinction in our model SED fit of $A_J=2$.

\indent The 92.6 day periodic photometric variability cannot be associated with the Keplerian orbit of the projected $\sim$40 AU visual companion.  Invoking the discussion in \citet[]{pla08a}, the periodic photometric variability of YLW 16A is not readily attributable to starspots, chaotic disk extinction from accretion, nor other stellar activity induced variations that tend to operate on time-scales of less than a week.  Instead, \textit{we postulate that the long-term periodic photometric variability of YLW 16A indicates the presence of a tertiary companion of unknown mass within a few AU of one of the visual components and with an orbital period of 92.6 days.}  If the tertiary companion is approximately the same mass/brightness of the other two companions, and it is periodically eclipsed by a circum-binary disk, this scenario could explain both the periodic variability as well as the $\sim$1/3 reduction in flux between bright and faint states.  The lack of a detection of binarity in \citet[]{covey,dop05} indicates that this tertiary companion may instead be at a much smaller observed luminosity.  The SED model indicates that the luminosity of the hot dust component must also change substantially to explain the observed variability at IRAC mid-IR wavelengths.  In other words, at $K_s$ and IRAC bands, we are seeing possible periodic eclipses (shadowing) of some of the hot dust material in the system associated with this tertiary companion, rather than the proto-star photosphere itself.  This in turn implies a strong star-disk dynamical interaction.  

\indent The literature photometry for YLW 16A provides an indication of the stability of the system, supporting a Keplerian origin to produce the observed periodic variations.  $K_s$ band photometry from \citet{sim87} and \citet{wil89}, as given in Table 1, indicates that the bright state magnitude of $\sim$9.8 may have been consistent for over twenty years. However, the faint state value of $\sim$8.8 mag does not agree with our observations. This could be due to long-term evolution in primordial disk structure.   Follow-up observations with the \textit{Spitzer} Space Telescope and the YSOVAR program will further constrain the long-term stability of the observed periodic variability  \citep[]{morales11}.

Finally, the discovery of a system similar to WL 4 indicates that such systems may be common, and more may be uncovered with long-term photometric NIR and mid-IR intense photometric monitoring.  Both WL 4 and YLW 16A possess visual companions at wide separations.  This implies that a wide companion offers a plausible, if not required, mechanism to explain how a circum-binary disk could be warped with respect to the orbit of a hypothesized inner binary, to produce the observed periodic disk extinction.  These two systems join KH-15D in NGC 2264 and CHS 7797 in the Orion star-forming region \citep[]{herbst,rod1,rod2}.

\section{Conclusion and Future Work}

We identify 92.6 day periodic photometric variability for the YSO YLW 16A.  We confirm the system is also a visual binary with a projected separation of $\sim$40 AU.  We infer a possible triple system for YLW 16A, similar to the model proposed for WL 4 by \citet{pla08a}, indicating such systems may be common.  

The nature of the companion producing the observed periodic photometric variations is unknown.  High S/N near-infrared spectroscopic monitoring of the 2.3 micron CO feature for radial velocity variations over three months, especially during the bright state, may confirm the presence of a tertiary companion responsible for producing the observed photometric variations.  Synoptic 3.6 and 4.5 $\mu$m observations have been obtained for both WL 4 and YLW 16A as part of the YSOVAR \textit{Spitzer} program \citep[]{morales11}.  Preliminary analysis indicates that the sources retain the photometric variability expected if the variability is driven by a Keplerian companion.  The analysis of this data will be reported in a future publication.

The authors would like to thank the anonymous referee for the manuscripts review.  We thank Karl Stapelfeldt, John Stauffer, Lynne Hillebrand and Andreas Seifahrt for their critical comments and discussion.  This research has made use of the NASA Exoplanet Archive, which is operated by the California Institute of Technology, under contract with the National Aeronautics and Space Adminstration under the Exoplanet Exploration Program.  This work is based (in part) on observations made with the Spitzer Space Telescope, which is operated by the Jet Propulsion Laboratory, California Institute of Technology under a contract with NASA.  Support for this work was provided by NASA through an award issued by JPL/Caltech.  This research has made use of the NASA/IPAC Infrared Science Archive, which is operated by the Jet Propulsion Laboratory, California Institute of Technology, under contract with the National Aeronautics and Space Administration.

\clearpage

\clearpage
\begin{figure}[ht] 
\begin{center}
\includegraphics[width=0.75\textwidth, clip=true,trim=0cm 0cm 0cm 0cm]{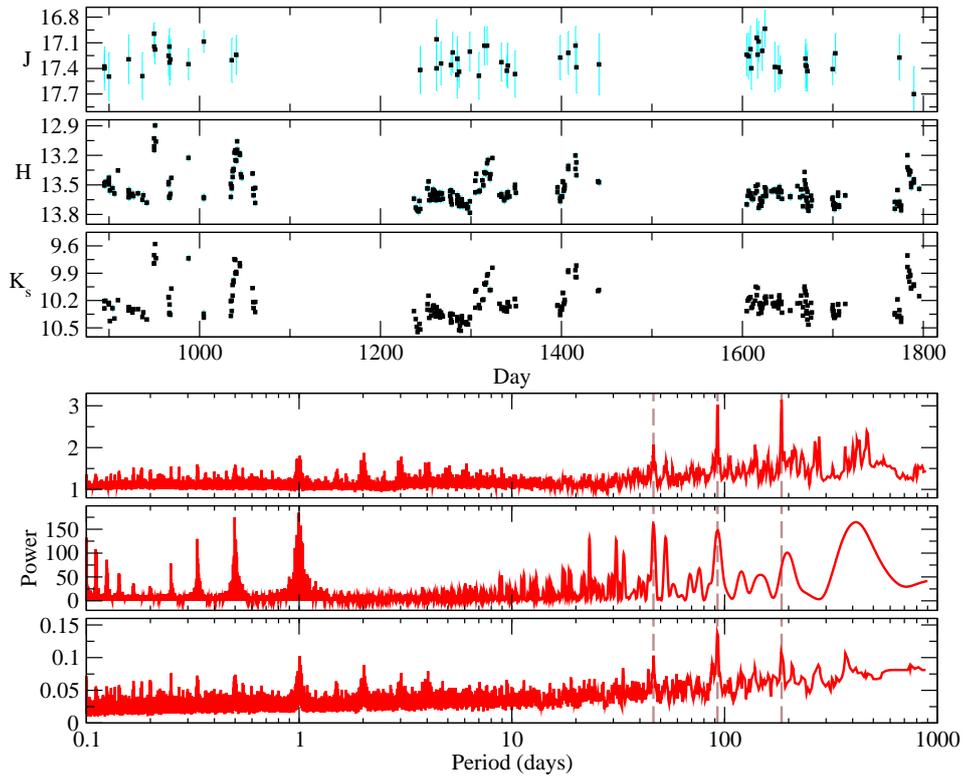}
\end{center}
\caption{Top three panels: the $J$, $H$, \& $K_s$-band light curves for YLW 16A, generated using data from the 2MASS Cal-PSWDB. ``Scan groups'' of six measurements taken in 10 minutes of elapsed real time are co-added, as in \citet[]{pla08b}.  Propagated 1-$\sigma$ error bars are shown in teal. Bottom three panels: Plavchan, Lomb-Scargle and Box Least Squares periodograms of the $K_s$-band light curve \citep[]{sca82,bls,pla08b}  } 
\label{fig:UnphasedCurve}
\end{figure}

\clearpage
\begin{figure}[ht] 
\begin{center}
\includegraphics[width=0.75\textwidth]{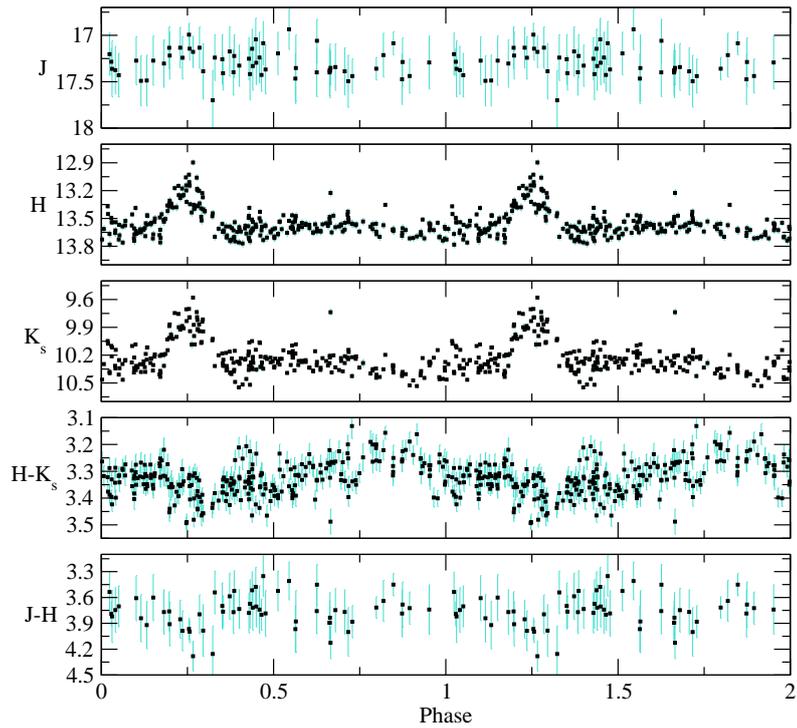}
\end{center}
\caption{From top to bottom: the $J$, $H$, $K_s$, $H$-$K_s$, \& $J$-$H$ phased light and color for YLW 16A, generated using data from the 2MASS Cal-PSWDB as in Figure 2, folded to a period of 92.6 days and plotted as a function of phase. A second phase of the same data is repeated.} 
\label{fig:ColorCurves}
\end{figure}

\clearpage
\begin{figure}[ht] 
\begin{center} 
\includegraphics[width=0.9\textwidth]{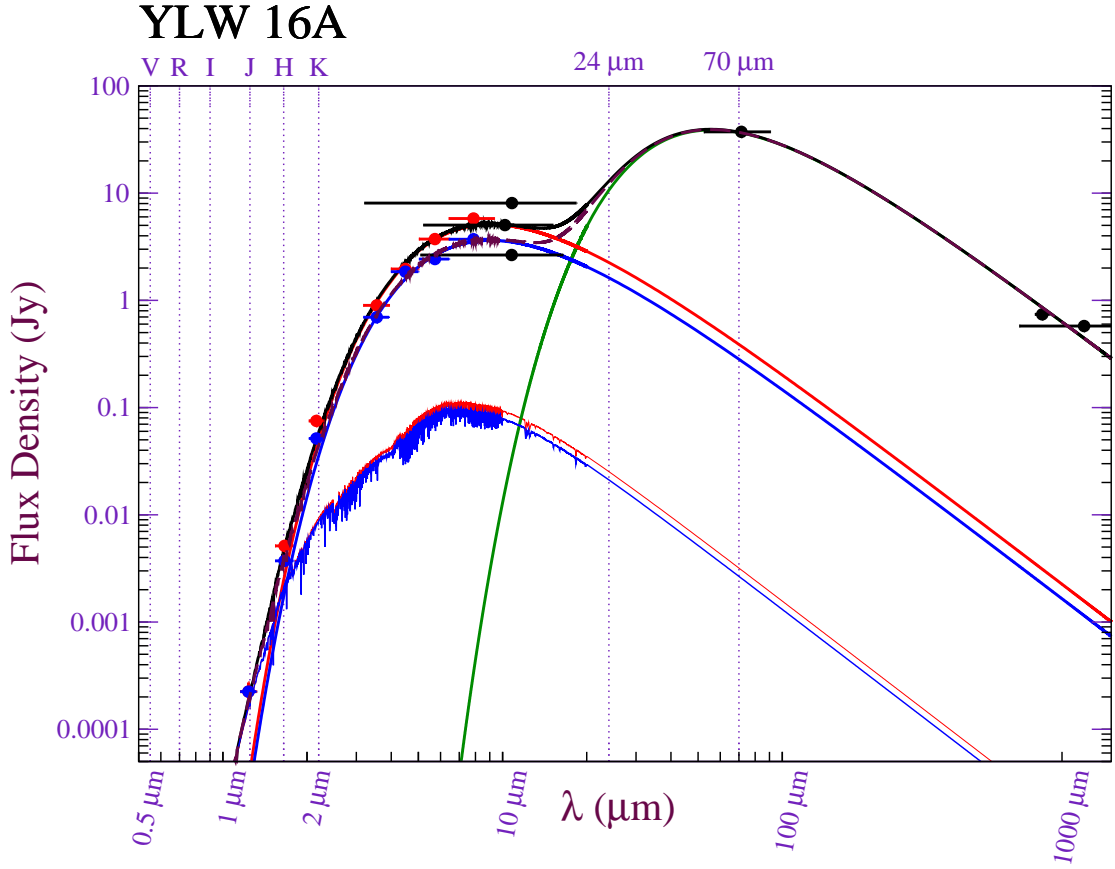}
\end{center}
\caption{The model SED fit to observed YLW 16A photometry. Blue circles correspond to $JHK$ and \textit{Spitzer} IRAC photometry during the bright state; red circles correspond to the faint state.  The summed model SED fit (black curve: bright state; maroon dashed curve: faint state) has contributions from a composite star (lower synthetic SEDs peaking at $\sim$0.1 Jy; red: bright state; blue: faint state), hot dust (upper left blackbody curves peaking at a few Jy, red: bright state; blue: faint state), and cold dust (green blackbody curve).  Ground-based historical 10.8 $\mu$m photometry, \textit{Spitzer} MIPS 70 micron photometry and sub-mm photometry is shown as black circles (YLW 16A is saturated at 24 $\mu$m in the \textit{Spitzer} c2d survey data).} 
\label{fig:SEDfit}
\end{figure}

\begin{landscape}

\clearpage
\begin{table}
\caption{YLW 16A Photometry}
\begin{tabular}{lllllllllllll}
$J$ & $H$ & $K_s$ & 3.6 $\mu$m$^a$ & 4.5 $\mu$m$^a$ & 5.8 $\mu$m$^a$ & 8 $\mu$m$^a$ & 10.8 $\mu$m & 70 $\mu$m$^e$ & 850 $\mu$m$^f$ & 1200 $\mu$m$^g$ & MJD & Phase$^h$\\
mag & mag & mag & mJy & mJy & mJy & mJy & mJy & mJy & mJy & mJy & days &  \\
 & & & 897$\pm$66 & 1960$\pm$154 & 3730$\pm$182 & 5800$\pm$350 &  & & &  & 53071.326 & 0.17 \\
 & & & 696$\pm$70 & 1850$\pm$168 & 2430$\pm$208 & 3760$\pm$345 &  & & &  & 53092.242 & 0.39  \\
 & & & & & & & & & & 576 & 52464.5 & 0.62\\
 & & & & & & & & & 740$\pm$148 & & &  \\
 & & & & & & & & 34700$\pm$3250 & & & & \\
 & & & & & & & 2650$\pm$65$^b$ & & & & 50626 & 0.77 \\
 & & & & & & & 8080$\pm$91$^c$ & & & & &   \\
 & & & & & & & (5030$\pm$460)$^d$ & & & & 46200 or & 0.98 or\\ 
 & & & & & & &  & & & & 46596 & 0.26\\
 & & 9.84$\pm$0.05$^i$ & & & & & & & & & 46596 & 0.26\\
 & & $\sim$8.8$^j$ & & & & & & & & & 46437 & 0.54 \\
17.13$\pm$0.1 & 13.25$\pm$0.1 & 9.87$\pm$0.1 & & & & & & & & & & bright state\\
17.13$\pm$0.1 & 13.60$\pm$0.1 & 10.28$\pm$0.1 & & & & & & & & & & faint state\\
 & & 42.2,9.3 mJy$^k$ & & & & & & & & & 53490 & 0.69\\ 
 & & & 351,344$^l$ & & & & & & & & 53469 & 0.46\\
 & & &  & 1209,641$^m$ & & & & & & & 54586.39 & 0.53\\
 
\end{tabular}
a Observations from the IRAC instrument \citep{eva03}. \\
b Observation made with P200/MIRLIN on 27 June 1997 \citep{bar05}.\\  This photometry is excluded from the SED fit optimization.\\
c Flux interpolated from 6.7 and 14.3 $\mu$m ISOCAM channels (data from \citet{bon01}). Observation date not given \citep{bar05}.\\  This photometry is excluded from the SED fit optimization.\\
d From Table 2 of \citet{wil89}; measured flux at 10.2 $\mu$m. Date uncertain; both May 1985 and June 1986 observing runs were cited in the text. \\This photometry is excluded from the SED fit optimization.\\
e Observation from the MIPS instrument \citep{pad08}.\\
f Observation details from \citet{joh04}. Error given as $\sim$20\% \citep{jor08}.\\
g Obtained between 7-12 July 2002 at the SIMBA bolometer array at the SEST telescope on La Silla, Chile \citep{sta06}. No error value given.\\
h See $\S$3. YLW 16A was observed to exhibit periodic variability. The light curve was folded such that the bright state corresponds to a phase of $\sim$0.17 to $\sim$0.3.\\
i Observation date given as June 1986; the 15th of June assumed \citep{wil89}. This photometry is excluded from the SED fit optimization.\\
j 7 January 1986. No error value given \citep{sim87}. This photometry is excluded from the SED fit optimization.\\
k NACO image fluxes of the left and right components in Figure 1a, scaled to the total flux from the 2MASS measurement in the faint state for $K_s$.\\
l NACO image fluxes  of the left and right components in Figure 1b, scaled to the total flux from the Spitzer IRAC 3.6$mu$m measurement in the faint state for $L$.\\
l CRIRES $M$-band fluxes  of the left and right components in Figure 1b, scaled to the total flux from the Spitzer IRAC 4.5$mu$m measurement in the faint state for $M$, from \citet[]{herczeg}.
\end{table}
 \clearpage
 \end{landscape}

\clearpage
\begin{table}
\caption{SED Model Parameters}

\begin{tabular}{ll}

Parameter & Value \\
Fixed & \\
\hline
Distance & 135 pc  \\
short-$\lambda$ extinction power law $\alpha$ & --2.0$^a$ \\
long-$\lambda$ extinction power law $\alpha$ & --1.0$^a$ \\
\hline
\multicolumn{2}{c}{Varying in faint state fit, fixed in bright state fit to faint fit} \\
\hline
$\lambda$ Extinction transition & 4.0 $\mu$m$^a$  \\
cold dust $T$, $L$ & 91.3 K, 1.92 L$_\odot$$^b$\\
cold dust $A_J$ & 4.2 mag $^b$\\
\hline
\multicolumn{2}{c}{Varying, Best Fit, Faint State} \\
\hline
composite $T_*$, $R_*$, $L_*$  & 3514 K, 4.27 $R_\odot$, 2.49 $L_\odot$  \\
$A_J$ stellar extinction & 9.86 mag\\
hot dust $T$, $L$ &  562 K, 1.10 L$_\odot$\\
A$_J$ hot dust &  0 mag$^c$ \\
\hline
\multicolumn{2}{c}{Varying, Best Fit, Bright State} \\
\hline
composite $T_*$, $R_*$, $L_*$  & 3525 K, 4.66 $R_\odot$, 3.01 $L_\odot$  \\
$A_J$ stellar extinction & 10.1 mag\\
hot dust $T$, $L$ &  579 K, 1.63 L$_\odot$\\
A$_J$ hot dust &  0.57 mag\\
\end{tabular}
\end{table}
a Extinction wavelength dependence adopted from \citet{bec78}; \citet{mat90}. A transition-wavelength initial guess of 3.5 $\mu$m was used in the fit, but was allowed to float freely in the fit to the faint state photometry.  The fit for the bright state photometry fixed the transition-wavelength to that of the best fit in the faint state.\\
b When we allow the cold dust temperature and luminosity to vary in our SED fit to the bright state photometry, the effect on the best fit parameters is marginal: $<$1 K, $<$1\% luminosity, ~$\sim$0.1 mag A$_J$ extinction difference.  Thus, we fix these values in the bright state fit to reduce the degrees of freedom.  The A$_J$ value for the cold dust is larger than the best fit hot dust extinction, which seems counterintuitive. This may relate to non-blackbody thermal dust grain emission, or the cold and hot dust being associated with different components of the visual binary.
c Negative extinction magnitudes were not allowed in the fit.


\newpage
\begin{thebibliography}{}

\bibitem[Adams et al.(1987)]{ada87} Adams, F. C., Lada, C. J., \& Shu, F. H.  1987, \apj, 312, 788
\bibitem[Allen et al.(2002)]{all02} Allen, L. E., Myers, P. C., Di Francesco, J., Mathieu, R., Chen, H., \& Young, E. 2002, \apj, 566, 993
\bibitem[Barsony et al.(2005)]{bar05} Barsony, M., Ressler, M. E., \& Marsh, K. A. 2005, \apj, 630, 381
\bibitem[Beckford et al.(2008)]{bec08} Beckford, A. F., Lucas, P. W., Chrysostomou, A. C., \& Gledhill, T. M.  2008, \mnras, 384, 908
\bibitem[Becklin et al.(1978)]{bec78} Becklin, E., Matthews, K., Neugebauer, G., \& Willner, S. 1978, \apj, 220, 831
\bibitem[Bertone et al.(2004)]{ber04} Bertone, E., Buzzoni, A., Chavez, M., \& Rodriguez-Merino, L. H.  2004, \aj, 128, 829
\bibitem[Bontemps et al.(2001)]{bon01} Bontemps, S., et al. 2001, \aap, 372, 173
\bibitem[Covey et al.(2006)]{covey} Covey, K., et al., 2006, AJ, 131, 512
\bibitem[Covey et al.(2013)]{coveypriv} Covey, K., 2013, private communication
\bibitem[Doppmann et al.(2005)]{dop05} Doppmann, G. W., Greene, T. P., Covey, K. R., \& Lada, C. J. 2005, \aj, 130, 1145 
\bibitem[Doyle et al.(2011)]{carter} Doyle, L., et al., 2011, Science,  333, 1602
\bibitem[Evans et al.(2003)]{eva03} Evans, N. J. II, et al. 2003, \pasp, 115, 965
\bibitem[Evans et al.(2009)]{evans09} Evans, N.J. II, et al., 2009, ApJS, 181, 321
\bibitem[Faesi et al.(2012)]{faesi}Faesi, C., et al., 2012, PASP, 124, 1137
\bibitem[Flaherty et al.(2012)]{flaherty} Flaherty, K. M., et al., 2012, ApJ, 748, 71
\bibitem[Flaherty \& Muzerolle(2010)]{flaherty2} Flaherty, K. M., \& Muzerolle, J., 2010, ApJ, 719, 1733
\bibitem[Greene \& Lada(2000)]{gre00} Greene, T. P., \& Lada, C. J.  2000, \aj, 120, 430
\bibitem[Girart et al.(2004)]{gir04} Girart, J. M., Curiel, S., Rodriguez, L. F., Honda, M., Canto, J., Okamoto, Y. K., \& Sako, S. 2004, \aj, 127, 2969
\bibitem[Grosso(2001)]{gro01} Grosso, N. 2001, \aap, 370, L22
\bibitem[Hauschildt et al.(1999)]{hau99} Hauschidlt, P., Allard, F., \& Baron, E.  1999, \apj, 512, 377
\bibitem[Herbst et al.(2010)]{herbst} Herbst, W., et al., 2010, AJ, 140, 2025
\bibitem[Herczeg et al.(2011)]{herczeg} Herczeg, G., et al., 2011, A\&A, 533, 112
\bibitem[Ida \& Lin(2010)]{planetformref2} Ida, S., \& Lin, D.N.C., 2010, ApJ, 719, 810
\bibitem[Imanishi et al.(2001)]{ima01} Imanishi, K., Koyama, K., \& Tsuboi, Y. 2001, \apj, 557, 747
\bibitem[Johnstone et al.(2004)]{joh04} Johnstone, D., Di Francesco, J., \& Kirk, H. 2004, \apj, 611, L45
\bibitem[Jorgensen et al.(2008)]{jor08} J$\o$rgensen, J. K., Johnstone, D., Kirk, H., Myers, P. C., Allen, L. E., \& Shirley, Y. L. 2008, \apj, 683, 822
\bibitem[Joy(1945)]{joy42} Joy, A. 1945, ApJ 102, 168 
\bibitem[Kovacs et al.(2002)]{bls} Kovacs, G., Zucker, S. \& Mazeh, T. 2002, A\&A, 391, 369
\bibitem[Lenzen et al.(2003)]{len03} Lenzen, R. Hartung, M., Brandner, W., Finger, G., Hubin, N. N., Lacombe, F., Lagrange, A.-M., Lehnert, M. D., Moorwood, A. F. M., \& Mouillet, D. 2003, SPIE, 4841, 944
\bibitem[Leous et al.(1991)]{leo91} Leous, J. A., Feigelson, E. D., Andre, P., \& Montmerle, T. 1991, \apj, 379, 683
\bibitem[Lin \& Papaloizou et al.(1980)]{planetformref1} Lin, D.N.C., \& Papaloizou, J, 1980, MNRAS, 191, 37
\bibitem[Lucas \& Roche(1998)]{luc98} Lucas, P. W., \& Roche, P. F. 1998, \mnras, 299, 699
\bibitem[Luhman \& Rieke(1999)]{luh99} Luhman, K. L., \& Rieke, G. H. 1999, \apj, 525, 440
\bibitem[Mamajek(2008)]{mam08} Mamajek, E. E.  2008, Astronomische Nachrichten, 329, 10
\bibitem[Marsh et al.(2010)]{marsh} Marsh, K.A.., 2010, ApJ, 719, 550
\bibitem[Mathis(1990)]{mat90} Mathis, J. S. 1990, \araa, 28, 37
\bibitem[Morales-Calderon et al.(2011)]{morales11} Morales-Calderon, M., et al., 2011, \apj, 733, 50
\bibitem[Nelder \& Mead (1965)]{metcalf} Nelder, J. A., \& Mead., R., 1965, Computer Journal, 7:308-313
\bibitem[Padgett et al.(2008)]{pad08} Padgett, D. L., et al. 2008, \apj, 672, 1013
\bibitem[Parks et al.(2013)]{par10} Parks, J. R., Plavchan, P., White, R., \& Gee, A. H., 2013, ApJS, submitted
\bibitem[Plavchan et al.(2008a)]{pla08a} Plavchan, P., Gee, A. H., Staplefeldt, K., \& Becker, A. 2008a, \apj, 684, L37
\bibitem[Plavchan et al.(2008b)]{pla08b} Plavchan, P., Jura, M., Kirkpatrick, J. D., Curti, R. M., \& Gallagher, S. C. 2008b, \apjs, 175, 191
\bibitem[Rebull(2001)]{rebull01} Rebull, L., 2001, AJ, 121, 1676 
\bibitem[Rodriguez-Ledesma et al.(2013)]{rod1} Rodriguez-Ledesma, M. V., et al., 2013, A\&A, 551, 44
\bibitem[Rodriguez-Ledesma et al.(2012)]{rod2} Rodriguez-Ledesma, M. V., et al., 2012, A\&A, 544, 112
\bibitem[Rousset et al.(2003)]{rou03} Rousset, G., et al. 2003, SPIE, 4839, 140
\bibitem[Scargle(1982)]{sca82} Scargle, J. D. 1982, \apj, 263, 835
\bibitem[Simon et al.(1987)]{sim87} Simon, M., Howell, R. R., Longmore, A. J., Wilking, B. A., Peterson, D. M., \& Chen, W. P. 1987, \apj, 320, 344
\bibitem[Skrutskie et al.(2006)]{skr06} Skrutskie, M. F., et al. 2006, \aj, 131, 1163
\bibitem[Stanke et al.(2006)]{sta06} Stanke, T., Smith, M. D., Gredel, R., \& Khanzadyan, T. 2006, \aap, 447, 609
\bibitem[Wilking et al.(1989)]{wil89} Wilking, B. A., Lada, C. J., \& Young, E. T. \apj, 340, 823
\bibitem[Xiao et al.(2012)]{covey2} Xiao, H.Y., et al., 2012, ApJS, 202, 7
\end{thebibliography}
\end{document}